\documentstyle[aps,epsfig]{revtex}
\advance\topmargin by 2.0cm
\advance\oddsidemargin by 1.3cm
\advance\evensidemargin by 1.3cm

\input epsf

\begin{document}

\begin{center}
\Large
{\bf
Why the Bradley aberration cannot be used to measure absolute speeds.
A comment.
} 
\end{center}

\begin{center}
Klaus Kassner \\
Institut f\"ur Theoretische Physik,
Otto-von-Guericke-Universit\"at Magdeburg,
Postfach 4120,
D-39016 Magdeburg, Germany\\[0.1cm]

29 August 2001

\end{center}

In a recent article in this journal \cite{sardin01}, Sardin 
proposed to use 
the Bradley aberration of light
for the construction
of a speedometer capable of measuring absolute speeds.
The purpose of this comment is to show that the device would not work.

Stellar aberration appears to be independent of the velocity
of the star observed. 
This fact, even though explained long ago  \cite{eisner67},
has remained a source of continuing confusion.
After all, it seems difficult to reconcile with the relativity principle.
Indeed, if the arguments of  \cite{sardin01} could be upheld, they
would  lead to a rejection of this principle.
To clarify the issue a bit, I will discuss different concepts of aberration
and their relation to the Bradley aberration.

Roughly speaking, aberration is the difference between the
observed and ``true'' angular positions of a star, caused by the motion
of either the observer or the star. A problem with this definition is
that there is no such thing as a true position. Positions have been 
observer dependent since the days of Newton. 
In order to get a workable definition of {\em true aberration},
 a distinguished observer is needed determining the frame in 
which to measure the true position of a star. Clearly, the most natural
choice would be an observer {\em at rest} with respect to the star, because
she will always see it at the same angle. Her frame of reference is
the same as that of the star, so I will 
consider just the frames of observer and star.
It is then easy to calculate the
so-defined  true aberration for various situations. For simplicity,
the calculations will be done with the velocities of the star and the 
observer(s) parallel to a prescribed line taken as $x$ axis.

First assume  the star to move at velocity $-{\bf V}$ (in a certain frame
$\Sigma'$) and the observer to be at rest. Let a light ray emitted at
an angle $\theta$ (see Fig.~1, left) in the star's frame $\Sigma$
hit the observer's
eye. What is the observed angle $\theta'$?
\begin{figure}[h]
\epsfig{file=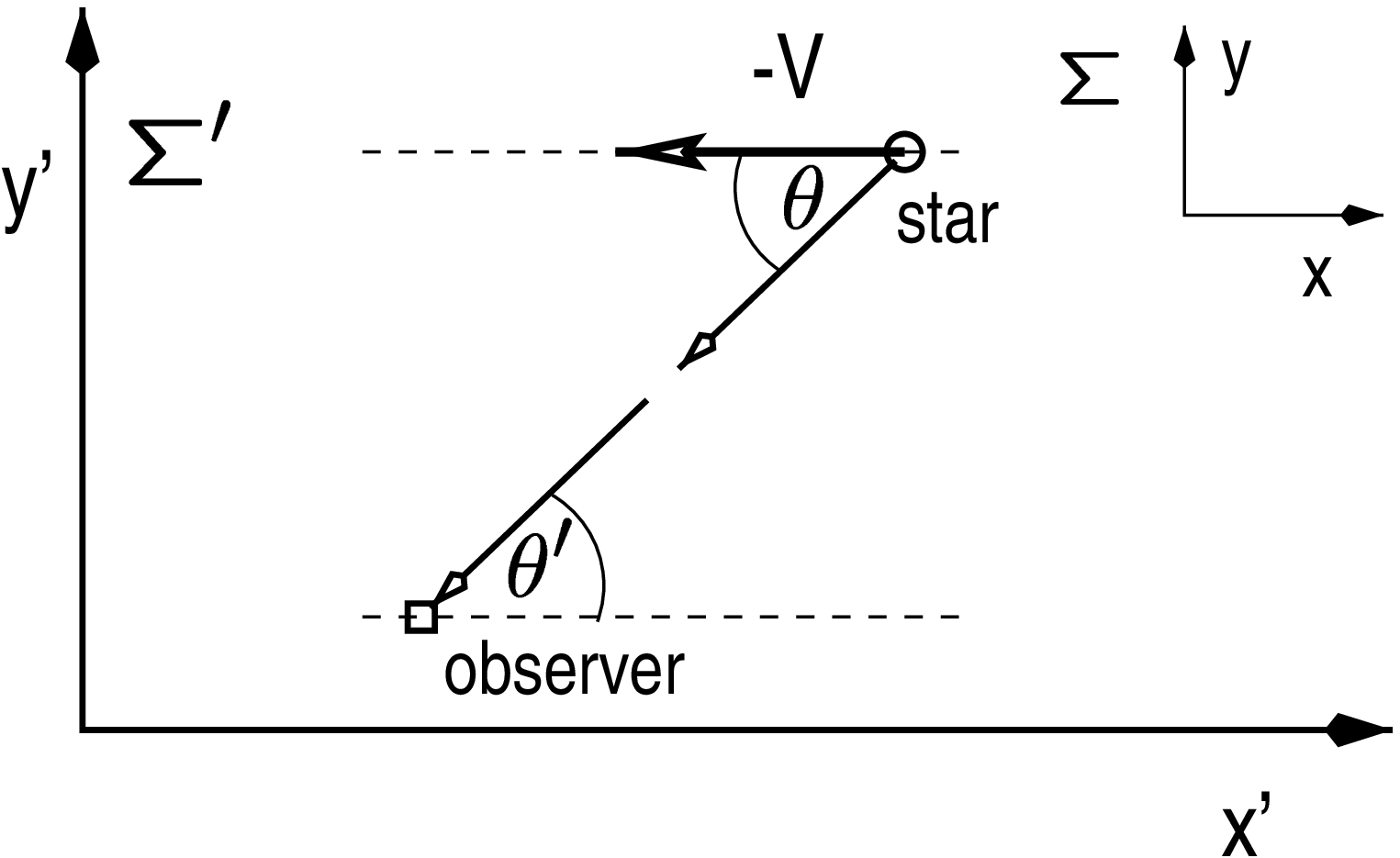,width=5.0cm,angle=0} \hspace{2.0cm}
\epsfig{file=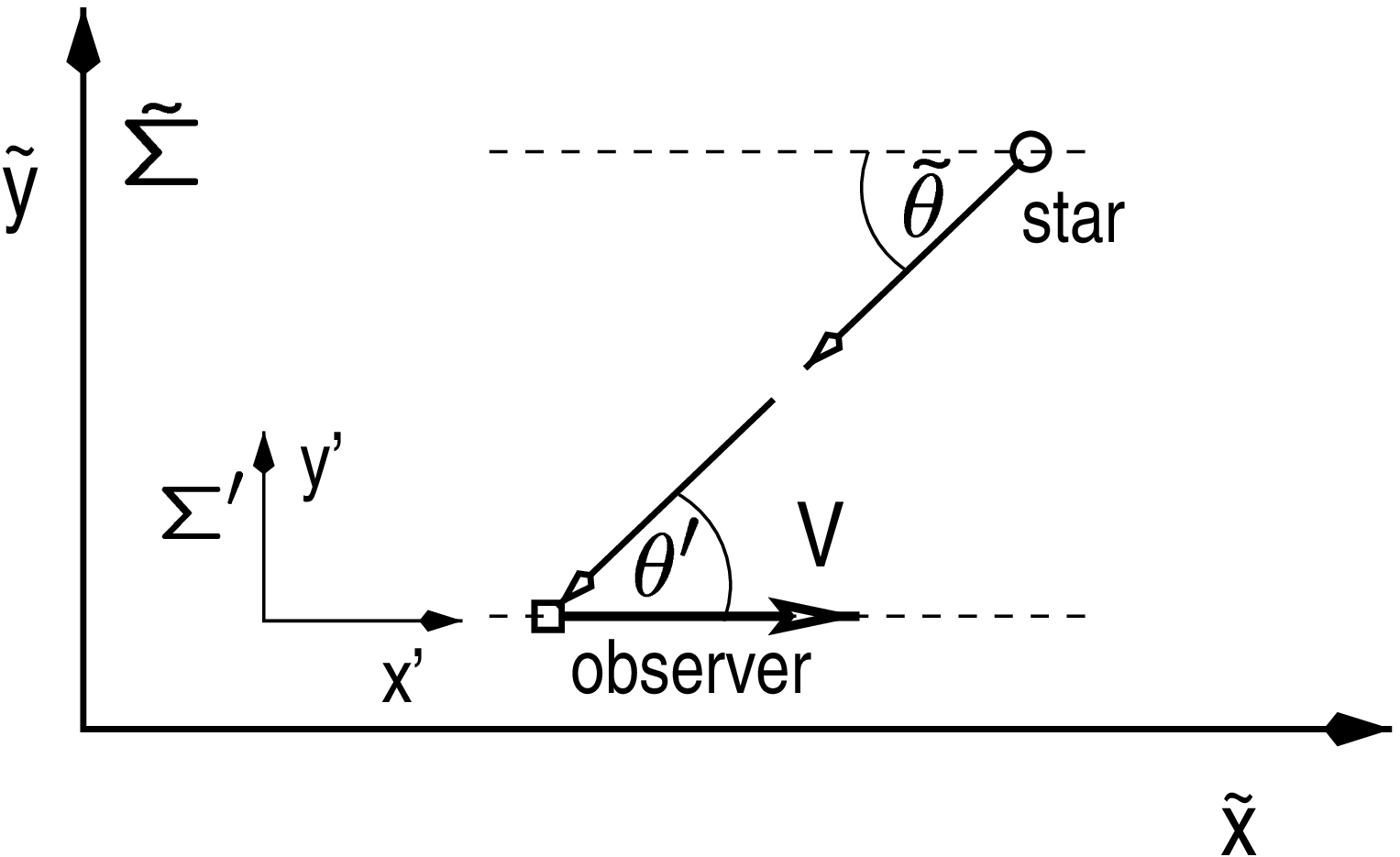,width=5.0cm,angle=0}
\caption{Aberration for moving source (left) 
and for moving observer (right).}
\label{figure1}
\end{figure}
The answer is provided by the relativistic addition theorem for velocities:
\begin{equation}
{\bf u'} = {\bf u'}_{\parallel} +{\bf u'}_{\perp} =
\frac{{\bf u}_{\parallel}-{\bf V}}{1-{\bf V u}/c^2} 
+ \frac{{\bf u}_{\perp}}{\gamma (1-{\bf V u}/c^2)} \>,
\label{veladd} 
\end{equation}
where ${\bf u}$ is the velocity in $\Sigma$, 
${\bf u}_{\parallel}\equiv {\bf V} ({\bf V u})/V^2$ and 
${\bf u}_{\perp}\equiv {\bf u}- {\bf u}_{\parallel}$
are its components parallel and perpendicular 
to ${\bf V}$, respectively,
 ${\bf u'}$, 
${\bf u'}_{\parallel}$, ${\bf u'}_{\perp}$ are the corresponding 
quantities in $\Sigma'$ and $\gamma = (1-V^2/c^2)^{-1/2}$.
Note that while the {\em speed} of light is the same in the two frames,
its  {\em velocity} (vector) is not.
In $\Sigma$, we have ${\bf u}=-c(\cos\theta {\bf e}_x +\sin\theta {\bf e}_y)$,
so we obtain  
${\bf u'}=-c(\cos\theta' {\bf e}_x +\sin\theta' {\bf e}_y)$,
with $\cos\theta'=(\cos\theta+V/c)/[1+(V/c)\cos\theta]$ and
$\sin\theta'=\sin\theta/[\gamma (1+(V/c)\cos\theta)]$, which can be
simplified, using $\tan x/2 = \sin x/(1+\cos x)$, to yield the well-known
result
\begin{equation}
\tan\frac{\theta'}{2} = \tan\frac{\theta}{2} \>\sqrt{\frac{c-V}{c+V}} \>.
\label{tanthethalf}
\end{equation}

Consider now the situation (Fig.~1, right) 
where the observer moves at ${\bf V}$ in some system
$\tilde\Sigma$, while the star is at rest (hence $\tilde\Sigma=\Sigma$). 
If the angle observed in $\Sigma'$ is  $\theta'$ again,
what will be its value in $\tilde\Sigma$?
One finds, using the addition theorem with ${\bf V}$ instead of $-{\bf V}$:
$\tan\tilde\theta/2 = \tan \theta'/2 \>\sqrt{(c+V)/(c-V)}$. Hence
$\tilde\theta=\theta$, and the {\em true aberration} is the same for the
two situations. It  depends on the {\em relative} velocities of
star and observer only. 

However, with an {\em a priori} unknown velocity of the star, 
an observer at rest with respect to it is not normally handy.
Therefore what is {\em measured} is not the true aberration, in general. 
Let us then define as {\em relative aberration} 
the difference in angles observed 
by two arbitrary observers moving at different
velocities ${\bf V_1}$ and ${\bf V_2}$ and looking at the same star
(the moment they meet). 
From (\ref{tanthethalf}), we have
$\tan \theta_1/2 = \tan \theta/2 \sqrt{(c-V_1^s)/(c+V_1^s)}$
and $\tan \theta_2/2 = \tan \theta/2 \sqrt{(c-V_2^s)/(c+V_2^s)}$,
where $V_1^s$ and $V_2^s$ are the velocities of the observers 
{\em relative} to the star. 
$\theta$ may be eliminated to obtain 
$\tan \theta_2/2= \tan \theta_1/2 
\sqrt{(c-V_2^s)(c+V_1^s)/(c+V_2^s)(c-V_1^s)}$. 
Using the velocity
addition theorem to express $V_{1/2}^s$ by $V_{1/2}$ and the
unknown velocity $W$ of the star, 
 $V_{1/2}^s=(V_{1/2}-W)/(1-V_{1/2}W/c^2)$,
we find 
$(c+V_{1/2}^s)/(c-V_{1/2}^s)=[(c-W)(c+V_{1/2})]/[(c+W)(c-V_{1/2})]$
and thus
\begin{equation}
\tan \frac{\theta_2}{2} = \tan \frac{\theta_1}{2}
\sqrt{\frac{(c-W)(c+V_1)}{(c+W)(c-V_1)}\frac{(c+W)(c-V_2)}{(c-W)(c+V_2)}}
= \tan \frac{\theta_1}{2}\sqrt{\frac{c-V_{21}}{c+V_{21}}} \>,
\label{tanthet12}
\end{equation}
where $V_{21} = (V_2-V_1)/(1-V_1V_2/c^2)$ is the {\em relative} velocity
between the {\em two observers}. The speed $W$ of the star cancels
out of the formula! Of course, we could have obtained this result directly
by using the same reasoning as in the derivation of (\ref{tanthethalf}).
But Eq.~(\ref{tanthet12}) is much more instructive, as it gives us also
{\em conditions} under which the velocity of the star drops out.

For what is measured as {\em Bradley aberration
is  not precisely the relative aberration}. We don't
simultaneously have two observers handy. What is measured is the 
difference in angles obtained by {\em one} observer in different states
of motion, at different times. But this is the same thing as 
the relative aberration {\em provided} the velocity of the star
is constant between measurements. Then the only velocity
that counts is $V_{21}$,
a {\em relative} velocity again.
{\em If} the velocity of the star changed, we would obtain, instead of
(\ref{tanthet12})
\begin{equation}
\tan \frac{\theta_2}{2} = \tan \frac{\theta_1}{2}
\sqrt{\frac{(c-W_1)(c+V_1)}{(c+W_1)(c-V_1)}
      \frac{(c+W_2)(c-V_2)}{(c-W_2)(c+V_2)}} \>,
\label{tanthet12mod}
\end{equation}
where now $W_1$ and $W_2$ are the velocities of the star, when
it emitted the light rays 
reaching the observer when she
had the velocities $V_1$ and $V_2$, respectively. 

Sardin's device to measure ``absolute speeds'' replaces the star by
a light source {\em moving along with the observer}.
Thus the velocity of the source {\em changes} between successive observations
at different observer speeds. Hence, the  independence of 
Bradley aberration of the source velocity does not hold.
But this is the central argument on which the working principle of
the device is based. Therefore, the device will not work. 
In fact, we can calculate the aberration measured by the 
device from (\ref{tanthet12mod}),
because  the velocities of
the light source are known here.
In any  state of uniform motion we have $W_1=V_1$, $W_2=V_2$.
The result $\tan\theta_2/2 =  \tan\theta_1/2$ indicates that there
will be no observable aberration. 
There is no need to abandon the relativity principle.



\begin{thebibliography}{10}
\bibitem{sardin01}
Sardin G., 
{\em Measure of the absolute speed through the
Bradley aberration of light beams on a three-axis frame},
{\sl Europhys. Lett.} {\bf 53} (2001) 310 .

\bibitem{eisner67}
Eisner E., {\em Aberration of Light from Binary Stars - a Paradox?},
{\sl Am. J. Phys.} {\bf 35} (1967) 817.



\end{thebibliography}
\end{document}